# Investigating the Impact of Age and Sex on Cataract Surgery Complications and Outcomes

Hadas Ben-Eli[1,2], Yaacov Cnaany[1], Itay Chowers[1], Ayelet Goldstein[3]

[1]*Department of Ophthalmology, Hadassah-Hebrew University Medical Center, Jerusalem, Israel*
[2]*Department of Optometry and Vision Science, Hadassah Academic College, Jerusalem, Israel*
[3]*Department of Computer Science, Hadassah Academic College, Jerusalem, Israel*

Corresponding author: Hadas Ben-Eli, POB 12000, Jerusalem, Israel, 91120. Tel: +972-50-4638006. E-mail: hadasben@hac.ac.il , hadasben@hadassah.org.il, https://orcid.org/0000-0002-2832-4221

**Running head:** Age and Sex Effects on Cataract Surgery

**Financial Support:** None

**The authors declare that they have no conflict of interest.**

What was known before:
- Prior studies demonstrated sex-based differences in visual outcomes post-cataract surgery, with women generally reporting poorer results, and age-related declines in postoperative visual acuity, especially for patients over 75.
- The role of resident surgeons in cataract surgery outcomes was recognized, with higher complication rates but comparable visual outcomes under close supervision, though sex-based differences under resident care were not extensively explored.

What this study adds:
- This study provides evidence of sex-specific age thresholds where postoperative BCVA improvements decline, identifying a significant post-surgery BCVA disparity between males and females that is not explained by differences in surgical complications or risk factor prevalence.
- The research introduces new considerations for preoperative evaluations and surgical planning, emphasizing the importance of sex and age in predicting outcomes and suggesting that surgical intervention timing may benefit from tailoring to these demographic factors.

**Abstract**

Background/Objectives - Cataract surgery, a very common and critical procedure for restoring vision, has outcomes that can vary based on patient demographics. This study aimed to elucidate the effects of age and sex on the risk factors, intraoperative complications, and postoperative outcomes of cataract surgery.

Subjects/Methods - Conducted as a single-center retrospective cohort study, it analyzed 691 eyes from 589 individuals who underwent surgery at a tertiary referral center, utilizing data from electronic medical records to assess preoperative risk factors, intraoperative complications, and pre- and post-operative best corrected visual acuity (BCVA) along with demographic data.

Results - The main results highlighted that males aged 65-75 years exhibited significantly higher rates of functional postoperative BCVA (91% for males vs. 79% for females, p=0.007), a disparity that is not explained by differences in surgical complications or risk factor prevalence. Furthermore, the study identified age-specific thresholds where BCVA improvements significantly declined beyond 65 years for females and 75 years for males. The likelihood of worsened BCVA post-surgery increased with age for both sexes, with a significant decline in BCVA improvement transitioning from 55-65 years to 65-75 years age groups.

Conclusions - The findings underscore the critical influence of both sex and age on cataract surgery outcomes, revealing significant sex-specific age thresholds that signal lesser improvements in postoperative BCVA. These insights advocate for the integration of patient age and sex into preoperative evaluations to better tailor the timing and planning of cataract surgery, ultimately aiming to optimize clinical outcomes.

1. **Introduction:**

Cataract affects roughly 95 million people worldwide and is the predominant cause of blindness in developing nations (Liu et al. 2017). The surgery, prevalent in advanced countries, substantially improves patients' quality of life and safety (Cullen KA, Hall MJ 2009; Hatch et al. 2012; Lamoureux et al. 2011).

Despite advancements, cataract surgery poses risks, influenced by patient profiles, surgeon expertise, and cataract type, potentially impacting outcomes. Identifying these factors is vital to enhance surgery success and safety.

The Cataract National Dataset (CND) reveals how factors such as age, sex, comorbidities, and surgeon skill correlate with complication risks like posterior capsule rupture (Narendran et al. 2009). It supports a risk assessment system predicting individual outcomes, considering factors including age, sex, diabetes, and UV exposure (McCarty et al. 1999; Lewis et al. 2004).

While men and women undergo cataract surgery at similar rates (Seah et al. 2002a), women report poorer visual function outcomes and postoperative care (McKee et al. 2005; Hughes et al. 2023; Hashemi et al. 2012; Quintana et al. 2013). It's uncertain if this gender disparity persists across age groups or if specific risk factors and intraoperative complications contribute to these results.

Increasing age correlates with worse refractive results and satisfaction after cataract surgery, with those over 75 less likely to see improvement (Seah et al. 2002a; McKee et al. 2005; Schein et al. 1995). It remains uncertain if these age-related declines affect men and women equally or at what age the decline is most pronounced (Hashemi et al. 2012a).

In cataract surgery, the surgeon's expertise and the level of supervision significantly affect complications and visual results. A large-scale study with 48,377 cases found that residents had higher complication rates than experienced surgeons but achieved comparable visual outcomes, due to varied case types and close supervision. Risk factors for poor visual results included age over 65 and specific complications like dropped nuclei or post-surgical retina, cornea, and lens issues (Ti et al. 2014).

Puri et al. explored how supervisory experience affects phacoemulsification outcomes when performed by residents. They found that residents supervised by less experienced surgeons encountered more issues, like vitreous loss and anterior capsule tear, indicating a learning curve in surgical skill development (Puri et al. 2015). However, the research didn't consider sex-based differences, suggesting an area for further study on sex's influence on surgical outcomes.

Bottom of Form

Given the projected rise in cataracts from an aging population, personalized medical and public health strategies addressing patient-specific demographics and risks are crucial for better surgical outcomes (McCarty et al. 1999).

This study examines how patient demographics, especially age and sex, and resident surgeon expertise affect cataract surgery outcomes. It focuses on the role of these demographics and other risk factors in surgery efficacy and complications under resident supervision. A key goal is to discern sex-based disparities in outcomes among different age groups, and whether sex impacts surgery success consistently or varies with age, thus offering insights into the interplay of demographic factors and surgical experience on cataract surgery success.

## 2. Methods

2.1 Study Population

This study used retrospective data from January 2018 to February 2022, from Hadassah Medical Center's ophthalmology department, considering patients with cataract surgeries by residents. Exclusions were for missing preoperative assessments or incomplete data.

In Israel, ophthalmology residents start a five-year program, engaging in surgeries from late in the first year. Complex or high-risk surgeries are allotted to seasoned residents or attendings.

The study received ethical approval from the institutional Helsinki Committee (HMO-0459-18), aligning with the Declaration of Helsinki and ensuring patient confidentiality.

Each surgery was treated as a separate event due to the potential differences in complications and outcomes between eyes of the same patient, which is critical for evaluating the impact of age and sex on surgical risks and results.

2.2 Risk Assessment in Participants

Before surgery, a risk assessment based on a 0-22 scoring system evaluated patient factors like age, axial length, and cataract severity, alongside conditions such as Diabetic Retinopathy and prior ocular procedures. Higher scores suggested greater complication risk. (Cnaany Y et al. 2024).

2.3 Monitoring Surgical Complications

Surgical reports were meticulously reviewed for complications using predefined search terms. Thirteen types were recorded, including capsule tears, vitreous loss, and lens dislocation, among others, noting their occurrence and timing during surgery.

2.4 Evaluation of Surgical Outcomes

Surgical outcomes were assessed by comparing pre- and post-operative best corrected visual acuity (BCVA, LogMAR units). The improvement rate was calculated using the formula: (pre-operative BCVA - post-operative BCVA) / pre-operative BCVA. Cases meeting Israeli driving vision standards (BCVA under 0.3 LogMAR) were highlighted. Also noted were cases without BCVA improvement. The study focused on 656 patients with complete one-year visual data to assess short-term outcomes.

*2.5 Statistical analysis*

The D'Agostino-Pearson test checked for normality; the Mann-Whitney and Student's T-tests analyzed non-normal and normal distributions, respectively. The Z proportion test compared categorical variables, and the Chi-Square Test determined sex differences across ages. Bonferroni's correction controlled for multiple comparisons, with significance set at p<0.05/number of comparisons. Cohort comparisons only included groups with n>30 for robust analysis. Age stratification was by decade, and Scipy in Python was used for analysis. [19].

## 3. Results

3.1 Characteristics of Patients

The study analyzed 691 eyes from 589 patients. Patients' mean age was 71.5± 10.7 years, with a range of 24-97 years, and 49% were female. Cataract surgeries increased with age, peaking at 65-75 years and decreasing after 75, especially beyond 85 years. The age distribution of surgeries, depicted in Figure A.1 in the supplementary material, showed no significant sex differences (p=0.47). Preoperative BCVA averaged 0.71±0.71 LogMAR, improving to 0.24±0.33 postoperatively—a 58.8±36.4%. mean increase.

*3.2* Risk Factors for Surgical Complications

Sex-based Prevalence of Risk Factors

Risk factor analysis, presented in Table 1, revealed no sex-based differences except for oral alpha-1 antagonists, significantly more common in males (23.9%) than females (1.2%; p<0.001). Females averaged fewer risk factors (0.59±0.79) than males (0.86±1.0; p<0.001) and had a lower preoperative cataract score (2.6±2.5 vs. 3.1±2.8; p=0.03).

Age-Related Risk Factor Analysis

Patients with small pupils, using oral alpha-1 antagonists, or with phacodonesis and dense cataract were older on average (74.9±9.7, 77.1±9.1, and 78.2±10.4 years, respectively) than those without these risk factors (70.7±11.0, 70.5±10.9, and 71.1±10.8 years, respectively; p<0.001). Conversely, patients with diabetic retinopathy were younger than those without it (66.7 vs. 71.7 years; p=0.001). Table 2 lists mean ages for groups with and without certain risk factors; no significant age differences were noted for shallow anterior chamber or poor patient cooperation. Stratified analysis showed age-related trends in risk factor prevalence, which are illustrated in Figure A.2 in the supplementary files. The prevalence of extreme axial length decreased from the under-55 age group to the 55-65 age group (from 8.6% to 0.9%; p=0.01), but this wasn't significant after Bonferroni adjustment. The prevalence of oral Alpha-1 antagonist use rose from the 55-65 to the 65-75 age group (from 3.6% to 11.6%; p=0.01), while diabetic retinopathy (DR) decreased (from 14.7% to 7.9%; p=0.05), neither remaining significant post-adjustment.

A marked increase in risk factors like phacodonesis (from 3.8% to 15.5%; p=0.001) and dense cataract (from 36.3% to 60.3%; p=0.001) was noted between the 75-85 to the above- 85 age group. DR prevalence declined with age. For a comprehensive risk factor comparison stratified by age and sex, refer to Figure A.3 in the supplementary documents.

*3.3 Surgery Complications*

Sex-based Prevalence of Complications

A sex-based comparison of post-surgery complication rates revealed no significant differences between male and female patients. Detailed analyses by age and sex, are presented in Table A.1 and Figure A.4 in the supplementary materials.

Influence of Patient's Age on Surgery Complications

Stratification by age showed varying complication rates; PC tear and vitreous loss differed across ages. Iris complication prevalence rose in patients over 85 compared to those 75-85 (6.9% vs. 1%; p=0.006), though this trend was not significant after Bonferroni correction. For visual data, please see Figure A.5 in the supplementary materials.

*3.4 Surgery Outcomes*

Effect of Patient Age on Surgical Outcomes

An age-stratified analysis assessed the impact on BCVA improvement (Figure 1). All ages saw gains, but the increase diminished significantly from 72% in those aged 55-65 to 61% in the 65-75 bracket, and down to 51% in the 75-85 range (p=0.002 and p=0.0004, respectively). Post-operative BCVA was significantly better in younger patients. The 75-85 age group had a worse average BCVA than the 65-75 group (p=0.01), and those over 85 had even higher (poorer) BCVA than the 75-85 group (p=0.03). Post-Bonferroni correction, these differences were not statistically significant.

Further analysis, with sex stratification shown in Figure 2, indicated significant age-related differences in visual acuity improvement post-surgery. Women over 65 showed a significant reduction in improvement, from 0.76±0.2 in the 55-65 age group to 0.58±0.2 in the 65-75 group (p=0.003). For men, a significant decrease in improvement was observed after age 75, from 0.64±0.3 in the 65-75 group to 0.52±0.3 in the 75-85 group (p=0.002). Overall, men had a higher improvement percentage in most age groups, except for the 55-65 age group, where women had better outcomes.

Further, the study found variations in BCVA before and after surgery. For females, pre-operative BCVA differed significantly between the 55-65 and 65-75 age groups (1.24±0.99 vs 0.73±0.75, p=0.009), and between the 75-85 and over 85 age groups (0.53±0.44 vs 0.84±0.64, p=0.006). For males, a significant difference in post-operative BCVA was noted between the 65-75 and 75-85 age groups (0.17±0.18 vs 0.24±0.28, p=0.005). In the under-55 group, pre-operative BCVA was notably lower in males than females (0.74±0.68 vs. 1.19±1.08, p=0.01). Given the small size of the male cohort (n=12), these results should be interpreted with caution.

 In-depth analysis showed more males had functional post-operative BCVA across age groups, significantly so in those aged 65-75 (males 91% vs females 79%, p=0.007), as shown in Figure 3.

Analyzing by sex alone, males had a significantly higher rate of functional post-operative BCVA than females (86% vs. 79%, p=0.01), with both genders showing the same rate of functional pre-operative BCVA (34%), as shown in Table A.2 in supplementary. This aligns with greater BCVA improvement in males aged 65-75 and above.

When examining changes in BCVA post-surgery by age, a clear pattern emerged: the likelihood of no improvement or worsening BCVA increased with patient age for both genders. Females had higher rates of these outcomes in every age category. A notable amount of females under 55 showed no BCVA improvement post-surgery, as detailed in Figure A.6 in the supplementary materials.

## 4. Discussion

In this study we evaluated the association of age and sex with risk factors, intraoperative complications, and postoperative BCVA outcomes in cataract surgery.

Sex-specific findings showed equal surgery rates across genders, consistent with previous studies [9]. Post-surgery, males typically had better outcomes. Notably, women over 65 had higher functional preoperative BCVA, whereas men were more likely to achieve superior functional postoperative BCVA, especially noticeable in the 65-75 age group. Men also showed more significant visual acuity improvements across most age groups, except for those 55-65, where women, starting from a higher preoperative BCVA, improved more.

The findings correspond with earlier studies showing females often have less favorable surgical outcomes [12,20] These disparities prompt further investigation into their consistency across age groups and possible links with pre-operative risks or intra-operative complications in females.

Our analysis didn't find notable sex-based differences in surgical complications, which contrasts with findings from Geiger et al. [21] and Morano et al. [22], that observed more postoperative complications in males. Our data did show males have a higher use of oral alpha-1 antagonists, which increases with age and is linked to intraoperative floppy iris syndrome (IFIS) (Christou et al. 2022; Blouin et al. 2007; Kaczmarek, Prost, and Wasyluk 2019; Karaca et al. 2021; Maluskova et al. 2023; Herranz-Cabarcos et al 2023). Given these drugs are also prescribed for conditions like COVID-19/SARS, Parkinson's disease, PTSD [29] and urinary symptoms in women[30–32] vigilance for IFIS is necessary during cataract surgery for both sexes.

Despite more males under 55 having risk factors like Oral-alpha and diabetic retinopathy, it didn't lead to more postoperative complications, highlighting the complexity of factors influencing surgical outcomes. The observed postoperative discrepancies between sexes, therefore, could not be explained by risk factors or complications, pointing to the potential influence of subjective aspects like BCVA assessment.

Studies have shown that women report lower visual acuity than men, a discrepancy potentially related to their lesser engagement in post-surgery care such as laser capsulotomy and refractive error correction, according to VF14 questionnaire results pre- and 4 months post-surgery.(McKee et al. 2005; Hashemi et al. 2012) Subjective measures and psychosocial factors could also play a role in shaping these sex-specific perceived outcomes after surgery. Thus, while our findings align with previous reports of males achieving better postoperative outcomes, the reasons for this advantage are not yet clear since they don't correspond to risk factors or complications. This emphasizes the need to factor in both physiological and psychological elements when assessing postoperative success across genders (McKee et al. 2005).

Most cataract surgery patients were 65-75 years old, despite literature indicating increased cataract occurrence with age.(Seah et al. 2002). This may suggest a shift towards surgeries for visual enhancement, which could also be influenced by our institutional practices where residents are less likely to perform surgeries on the very elderly (Cnaany Y et al. 2024).

An important finding from our research is the impact of age on cataract surgery results, especially in BCVA enhancement. We identified gender-related age thresholds where BCVA gains decline—post-65 for women and post-75 for men—which are critical considerations for surgical timing. While younger patients typically experience more significant enhancements in postoperative BCVA, a trend consistent with other research (Jaycock et al. 2009; Muhtaseb et al 2004), it's crucial to note that all age groups benefit from surgery. Patients over 85 still achieve notable improvement, averaging 50% better BCVA, with 65% reaching functional vision post-surgery for both genders. These outcomes argue against delaying surgery for the elderly due to significant life quality impacts and increased fall risks. [36] Even those over 96 have meaningful visual gains, although reduced compared to younger individuals, possibly due to more severe age-related maculopathy (Syam et al. 2004; Jaycock et al. 2009).

Moreover, our research identified age-related risk factors for cataract surgery that affect both genders. These include oral alpha-1 antagonists use, small pupil size, phacodonesis, and dense cataracts, all of which increase with age. These conditions are often linked to pseudoexfoliation (PXF), the accumulation of fibrillar material in ocular tissues, complicating surgery(Plateroti et al. 2015; Borkenstein and Borkenstein 2019;  Buhbut et al. 2023; Grzybowski et al. 2019).

Contrary to other risk factors, patients with diabetic retinopathy (DR) were notably younger than those without, suggesting diabetes may hasten cataract development, requiring sooner surgery. This aligns with studies designating DR as an indicator for early cataract surgery. (Tomić et al. 2021; Kiziltoprak et al. 2019).

Addressing the limitations of our study, the limited number of patients with specific complications may have influenced the statistical significance of age-related differences. Additionally, the retrospective design and potential EMR data limitations may have led to missing nuanced or unrecorded complications, such as intraoperative floppy iris syndrome. Despite these challenges, we implemented stringent data validation methods to enhance reliability. The focus on resident-performed surgeries and the single-center nature of the study could restrict the applicability of our findings to wider surgical practices and settings. Moreover, while our one-year follow-up conforms to common practices, it might not be sufficient to detect long-term complications or the persistence of visual acuity improvements. A more extended follow-up, inclusive of a wider array of surgeons across multiple institutions, could yield more comprehensive insights into the long-term effects of cataract surgery.

**Conclusions**

Our comprehensive analysis has underscored the critical roles of sex and age in determining the outcomes of cataract surgery. Notably, the distinct patterns observed in women's postoperative results underscore the inadequacy of relying solely on traditional clinical metrics. These patterns affirm the importance of incorporating a broader range of physiological and psychological factors into preoperative assessments and postoperative care plans to optimize surgical outcomes.

**Acknowledgments and Funding**

a. This research received no specific grant from any funding agency in the public, commercial, or not-for-profit sectors.

b. The authors express their gratitude to Dr. Ron Kaufman and Dr. Ori Saban for their valuable insights that enhanced this article.

**Funding:** The authors declare no financial disclosures in this study.

**Data Availability:** Data are available on reasonable request.

**Ethical Approval:** The study was approved and performed in accordance with the institutional Helsinki Committee (study #: HMO-0459-18). The committee exempts retrospective research from informed consent by the participants. Data was collected from the Ophthalmology Department database and anonymized before analysis. This study was performed in accordance with the Helsinki Declaration of 1964, and its later amendments.

**Author Contribution Statement:** All authors contributed to the study's conception and design. Material preparation, data collection, and analysis were performed by Ayelet Goldstein, Hadas Ben-Eli, Yaacov Cnaany, and Itay Chowers. The first draft of the manuscript was written by Ayelet Goldstein, Hadas Ben-Eli, and Yaacov Cnaany. All authors commented on previous versions of the manuscript. All authors read and approved the final manuscript.

**Titles and legends to figures**

**Figure 1. Comparative Analysis of Pre- and post-operative Best Corrected Visual Acuity (BCVA) and Percentage of Improvement Across Age Groups** The dual-axis graph presents Best Corrected Visual Acuity (BCVA) measured in LogMAR before (blue bars) and after (orange bars) cataract surgery, alongside the percentage of visual acuity improvement (red line), across five distinct age groups. The left vertical axis corresponds to the BCVA (LogMAR) values, with lower scores indicating better visual acuity. The right vertical axis denotes the percentage of BCVA improvement, with higher values representing greater visual recovery. Error bars indicate the standard deviation (SD) within each age group, providing a measure of variability around the mean BCVA scores. Statistical significance between consecutive age groups, following Bonferroni correction for multiple comparison, is marked by the respective p-value, illustrating the differential impact of surgery on visual outcomes by age.

**Figure 2. Stratified Visual Acuity Outcomes by Age Group and Sex Before and After Surgery.** Best Corrected Visual Acuity (BCVA) measured in LogMAR for males (blue and dark green bars for pre-operative and post-operative, respectively) and females (red and light green bars for pre-operative and post-operative, respectively) across different age groups. Additionally, the lines indicate the percentage of visual acuity improvement post-surgery for each sex (male in blue, female in red). The left vertical axis corresponds to BCVA (LogMAR) values, with lower scores indicating better visual acuity, while the right vertical axis represents the improvement percentage. Error bars indicate the standard deviation (SD) within each age group, providing a measure of variability around the mean BCVA scores. Statistically significant differences, post Bonferroni correction for multiple comparisons, between consecutive age groups are noted with p-values, showing how surgical outcomes and recovery rates vary with age and between sexes. 'M' denotes males, 'F' denotes females, and 'VA' stands for Visual Acuity.

**Figure 3. Sex-Based Comparison of Functional Pre- and Post-Operative BCVA Across Age Groups.** The left side is a comparison of the percentages of functional pre-operative BCVA between females and males across various age groups. The right side mirrors this comparison for post-operative BCVA, highlighting the significant sex differences in visual acuity outcomes. Point of statistical significance marked with the respective p-value.

**Tables:**

**Table 1 – Comparison of Risk Factors by Sex.** Significant differences are highlighted in bold.

| Risk factor | Female (n=339) | Male (n=352) | P* |
|---|---|---|---|
| Axial length <21.5 or >30 | 6 (1.8%) | 9 (2.6%) | 0.48 |
| Axial length 26-30 | 20 (5.9%) | 21 (5.9%) | 0.97 |
| Dense cataract | 129 (38.0%) | 134 (38.1%) | 0.99 |
| Only eye | 9 (2.6%) | 3 (0.8%) | 0.07 |
| Age >90 | 6 (1.8%) | 12 (3.4%) | 0.17 |
| Age 80-90 | 68 (20.0%) | 66 (18.7%) | 0.66 |
| Shallow anterior chamber | 16 (4.7%) | 11 (3.1%) | 0.28 |
| Small pupil | 44 (12.9%) | 60 (17.0%) | 0.13 |
| **Oral alpha-1 antagonists** | **4 (1.2%)** | **84 (23.9%)** | **<0.001** |
| Lens phacodonesis | 10 (2.9%) | 15 (4.3%) | 0.35 |
| Poor cooperation | 102 (30.1%) | 92 (26.1%) | 0.25 |
| Previous vitrectomy | 0 | 3 (0.8%) | 0.09 |
| Diabetic retinopathy | 19 (5.6%) | 33 (9.4%) | 0.06 |
| Fuchs' Endothelial Dystrophy | 6 (1.8%) | 4 (1.1%) | 0.48 |
| Any risk factor | 235 (69.3%) | 260 (73.9%) | 0.18 |
| **Risk factors SUM£** | **1.3±1.2** | **1.5±1.4** | **0.04** |
| **preScore£** | **2.6±2.5** | **3.1±2.8** | **0.04** |

\* Non-parametric Mann Whitney test was used for numerical variables (marked with £), and the Z-proportion test for categorical variables. NA=Not applicable. Statistically significant risk factors are highlighted in bold.

**Table 2 – Comparison of Mean Age Among Study Populations with and Without Specific Risk Factors and Surgery Complications.** Included are groups where n > 30, ensuring sufficient sample size for reliable statistical analysis. Significant differences highlighted in bold.

| Risk factor/Complication (group size with vs without) | RF/Complication Present (Mean age±SD, years) | RF/Complication Absent (Mean age ±SD, years) | P* |
|---|---|---|---|
| **Diabetic Retinopathy** | **66.7±8.6 (n=52)** | **71.7±11.0 (n=639)** | **0.0001** |
| Poor Cooperation | 71.5±13.3 (n=194) | 71.2±9.8 (n=497) | 0.06 |
| **Oral alpha-1 antagonists** | **77.1±9.1 (n=88)** | **70.5±10.9 (n=603)** | **<0.0001** |
| **Small pupil** | **74.9±9.7 (n=104)** | **70.7±11.0 (n=587)** | **<0.001** |
| Shallow anterior chamber | 74.8±10.7 (n=27) | 71.2±10.9 (n=664) | 0.09 |
| **Phacodonesis** | **78.2±10.4 (n=25)** | **71.1±10.8 (n=666)** | **<0.001** |
| **Dense cataract** | **72.2±12.3 (n=263)** | **70.8±9.9 (n=428)** | **0.009** |
| Axial length 26-30 (vs. normal 21.5-26) | 69.3±11.2 (n=41) | 71.6±10.7 (n=635) | 0.23 |
| **Any Risk factor vs none** | **72.8±11.4 (n=495)** | **67.5±8.3 (n=196)** | **<0.0001** |
| **Vitreous loss** | **72.2±12.3 (n=43)** | **71.2±10.9 (n=648)** | **0.009** |
| Posterior Chamber tear | 64.6±16.1 (n=64) | 71.1±10.8 (n=627) | 0.10 |
| Any Complication (vs. none) | 69.3±11.2 (n=91) | 70.8±9.9 (n=600) | 0.23 |

\* Analyzed by Mann Whitney test. Statistically significant differences between groups with and without each risk factor are denoted in bold text.